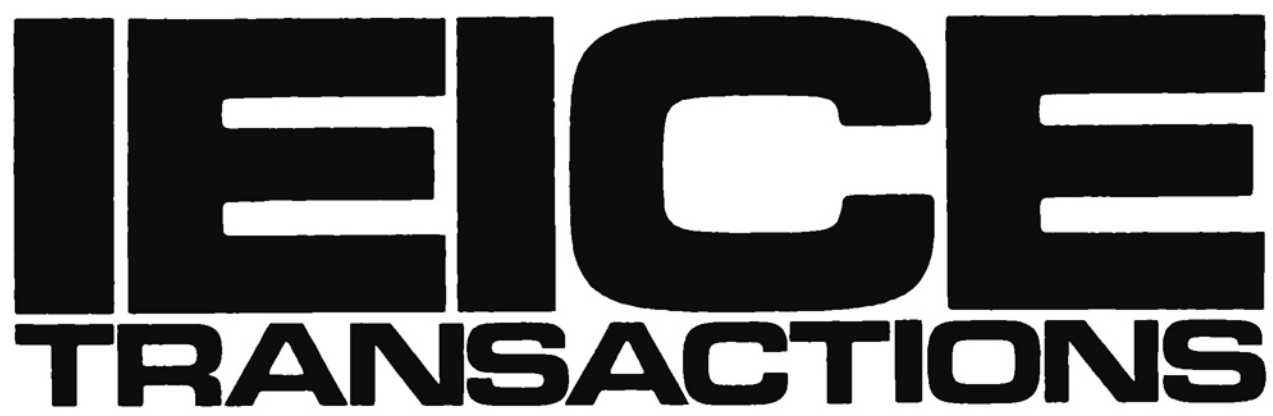

# on Information and Systems

PAPER

# Overlapped Wavelet Diffusion for Low-Light Image Enhancement

**Fen PENG**[†a)], ***Nonmember***, **Taizo SUZUKI**[††b)], ***and*** **Seisuke KYOCHI**[†††c)], ***Members***

**SUMMARY** In this study, we propose an overlapped wavelet diffusion framework for Low-Light Image Enhancement (LLIE), which incorporates two complementary components to achieve blocking artifact-free and detail-preserving enhancement. Although recent diffusion-based LLIE methods have demonstrated remarkable performance compared with traditional approaches, DiffLL still suffers from blocking artifacts caused by the Haar Wavelet Transform (WT) and blurred edges or over-smoothed textures due to the limitations of its High-Frequency Restoration Module (HFRM). To overcome these issues, we introduce an *Overlapped* WT (OWT) that incorporates correlations across neighboring regions, thereby structurally preventing blocking artifacts. Furthermore, we integrate a low-frequency-guided High-Frequency Enhance Block (HFEBlock) to strengthen detail recovery, yielding sharper edges and more reliable textures. Extensive experiments on the LOLv1 and LOLv2-real datasets demonstrate that our framework, termed "OWDiff," consistently outperforms existing LLIE methods both qualitatively and quantitatively, achieving superior visual quality while maintaining computational efficiency. OWDiff effectively addresses the structural limitations of the Haar WT and the HFRM, achieving an average PSNR gain of 0.58 dB, along with a 1.64% relative improvement in SSIM and a 5.9% relative reduction in LPIPS, compared to DiffLL across both the LOLv1 and LOLv2-real datasets.

## 1. Introduction

Low-light images often suffer from poor visibility, low contrast, and severe noise caused by environmental and hardware limitations. Consequently, such images lack aesthetic appeal and sufficient semantic information. The goal of Low-Light Image Enhancement (LLIE) is to improve brightness and contrast while preserving natural structures, which is essential for downstream tasks such as image classification [1], object detection [2], and autonomous driving [3].

Traditional LLIE methods, including histogram equalization [4–6], Retinex-based methods [7, 8], and gamma correction [9–11], can enhance brightness but often introduce color distortions or amplify noise. Deep learning approaches based on Convolutional Neural Networks (CNNs) [12–19] and Transformers [20, 21] have achieved stronger enhancement performance, yet still struggle with controlling enhancement strength and suppressing artifacts.

Diffusion-based LLIE models [22–24] have achieved high enhancement quality, but they remain sensitive to stochastic sampling and require high computational cost. DiffLL [23] improves efficiency by adopting the wavelet domain, where a Wavelet-based Conditional Diffusion Model (WCDM) enhances the low-frequency component and a High-Frequency Restoration Module (HFRM) attempts to recover details. However, DiffLL still suffers from two major limitations: (i) the 2D Haar Wavelet Transform (WT), a $2 \times 2$ block transform, introduces structural discontinuities that cause visible blocking artifacts; (ii) the HFRM is unable to adequately restore high-frequency structures buried in noise. In parallel, Mamba-based architectures have emerged as a promising new direction for LLIE, yet most existing models rely heavily on spatial-domain representations and degrade under extremely dark conditions where structural and frequency cues collapse. Among them, Wave-Mamba [25] stands out as the most relevant wavelet-domain Mamba model. It introduces a High-Frequency Enhance Block (HFEBlock) that reconstructs details without noticeable blocking artifacts, but its performance tends to deteriorate under extremely dark conditions due to the limited noise-handling ability of its 2D-Mamba backbone.

Motivated by these observations, we propose "OWDiff," which enhances DiffLL by integrating an *Overlapped* WT (OWT) and a wavelet-domain compatible HFEBlock. The OWT introduces structural continuity across neighboring regions, effectively suppressing blocking artifacts caused by non-overlapping transforms. Since OWT alone cannot fully recover high-frequency structures weakened in low-light images, we incorporate a low-frequency-guided HFEBlock to restore sharper edges and fine textures, resulting in a complementary design that balances structural continuity and high-frequency detail restoration. This is the first work to integrate a frequency-guided HFEBlock into a wavelet-domain diffusion framework, enabling effective frequency-guided refinement beyond prior diffusion-based LLIE methods. By combining OWT for structural continuity and HFEBlock for frequency-domain refinement, OWDiff achieves blocking artifact-free reconstruction and high-fidelity enhancement while retaining the computational efficiency of DiffLL. Despite potential concerns regarding error propagation in overlapped transforms, experiments on the LOLv1 [12] and the LOLv2-real [15] datasets demonstrate that OWDiff consistently outperforms existing methods both qualitatively and quantitatively.

[†]Master's Program in Computer Science, University of Tsukuba, Tsukuba-shi, 305-8573 Japan
[††]Institute of Systems and Information Engineering, University of Tsukuba, Tsukuba-shi, 305-8573 Japan
[†††]the Department of Digital Media, Hosei University, Koganei-shi, Japan
a) E-mail: peng@wmp.cs.tsukuba.ac.jp
b) E-mail: taizo@cs.tsukuba.ac.jp
c) E-mail: kyochi@hosei.ac.jp

Our key contributions are summarized as follows:

- Introducing an OWT that enforces cross-block structural continuity, eliminating blocking artifacts inherent to non-overlapping transforms.
- Incorporating a low-frequency-guided HFEBlock that recovers sharp edges and fine textures, compensating for the high-frequency attenuation introduced by OWT.

The integration of these components allows OWDiff to achieve blocking artifact-free and detail-preserving enhancement with high computational efficiency.

*Notation*: $\|\cdot\|_1$ and $\|\cdot\|_\mathrm{F}$ represent the $\ell_1$ and Frobenius norms, respectively.

## 2. Related Works

### 2.1 DiffLL

DiffLL [23] is a diffusion-based Low-Light Image Enhancement (LLIE) framework that incorporates the Wavelet Transform (WT) into the denoising process to improve both computational efficiency and robustness. Its pipeline consists of the following four stages:

1. Frequency decomposition via the Haar WT: A low-light image is decomposed into one low-frequency subband and three high-frequency subbands using the Haar WT, whose filter coefficients are listed in Table 1. This decomposition separates global brightness information from edges and textures, enabling frequency-wise processing.
2. Enhancement of the lowest-frequency subband using Wavelet-based Conditional Diffusion Model (WCDM): The lowest-frequency component is enhanced using the WCDM, which performs conditional diffusion-based denoising to improve brightness, contrast, and structural consistency.
3. Reconstruction of high-frequency subbands using High-Frequency Restoration Module (HFRM): The high-frequency subbands are processed using the HFRM, which aims to reconstruct edge and texture details from the decomposed representations.
4. Final image synthesis via inverse Haar WT: The enhanced lowest-frequency subband and the reconstructed high-frequency subbands are recursively fused using the inverse Haar WT to obtain the final enhanced image.

Despite its effectiveness, DiffLL still faces two major limitations:

- The Haar WT is a *block* transform that ignores cross-block correlations, often resulting in visible blocking artifacts during reconstruction.
- High-frequency reconstruction remains insufficient because fine details in low-light images are frequently dominated by noise, limiting the effectiveness of the HFRM.

### 2.2 Wave-Mamba

Wave-Mamba [25] is a wavelet-based Visual-State-Space Model (VSSM) designed for Ultra-High-Definition (UHD) LLIE. Its pipeline consists of the following four stages:

1. Initial feature extraction and wavelet decomposition: The process begins with a $3 \times 3$ convolution that extracts low-level feature embeddings, followed by decomposition into frequency subbands via the Haar WT. This step projects a UHD low-light image into the wavelet domain, enabling more efficient computation while preserving important structural information.
2. Low-frequency enhancement using Low-Frequency State Space Block (LFSSBlock): The low-frequency subbands are processed by the LFSSBlock, which leverages the long-range sequence modeling capability of State Space Models (SSMs) to improve brightness with linear computational complexity.
3. High-frequency refinement using High-Frequency Enhance Block (HFEBlock): The high-frequency subbands are refined using the HFEBlock, which employs the enhanced low-frequency features as guidance to correct and reconstruct high-frequency details through frequency matching. This design enables the recovery of sharper edges and more reliable textures without introducing blocking artifacts.
4. Progressive reconstruction via inverse Haar WT: The enhanced low-frequency and reconstructed high-frequency subbands at each wavelet level are progressively fused using the inverse Haar WT, producing a high-quality enhanced UHD image with improved brightness and structural fidelity.

Although Wave-Mamba effectively enhances UHD low-light images, its performance noticeably degrades under extremely dark conditions for two main reasons:

- Built upon a deterministic VSSM backbone, Wave-Mamba performs entirely input-driven reconstruction, utilizing the limited information available in severely dark regions, which often leads to artifacts or over-smoothing.
- Although Wave-Mamba restricts VSSM to low-frequency subbands to mitigate its inherent susceptibility to noise, residual noise in severely noise-overwhelmed regions can still propagate through its recurrent state updates, ultimately resulting in further noise amplification.

## 3. Overlapped Wavelet Diffusion

### 3.1 Motivation

OWDiff is designed to achieve LLIE without blocking artifacts while preserving fine details, a capability that neither

Table 1: Filter coefficients of WTs (low-pass: $\sqrt{2} \times \ell(n)$, high-pass: $1/\sqrt{2} \times h(n)$).

| | | $n=0$ | $n=1$ | $n=2$ | $n=3$ | $n=4$ | $n=5$ | $n=6$ | $n=7$ | $n=8$ |
|---|---|---|---|---|---|---|---|---|---|---|
| Haar WT | $\ell(n)$ | $1/2$ | $1/2$ | — | — | — | — | — | — | — |
| | $h(n)$ | $1$ | $-1$ | — | — | — | — | — | — | — |
| Bior2.2 WT | $\ell(n)$ | $-1/8$ | $1/4$ | $3/4$ | $1/4$ | $-1/8$ | — | — | — | — |
| | $h(n)$ | $1/2$ | $-1$ | $1/2$ | — | — | — | — | — | — |
| Bior4.4 WT | $\ell(n)$ | 0.0267 | −0.0169 | −0.0782 | 0.2669 | 0.6029 | 0.2669 | −0.0782 | −0.0169 | 0.0267 |
| | $h(n)$ | 0.0913 | −0.0575 | −0.5913 | 1.1151 | −0.5913 | −0.0575 | 0.0913 | — | — |

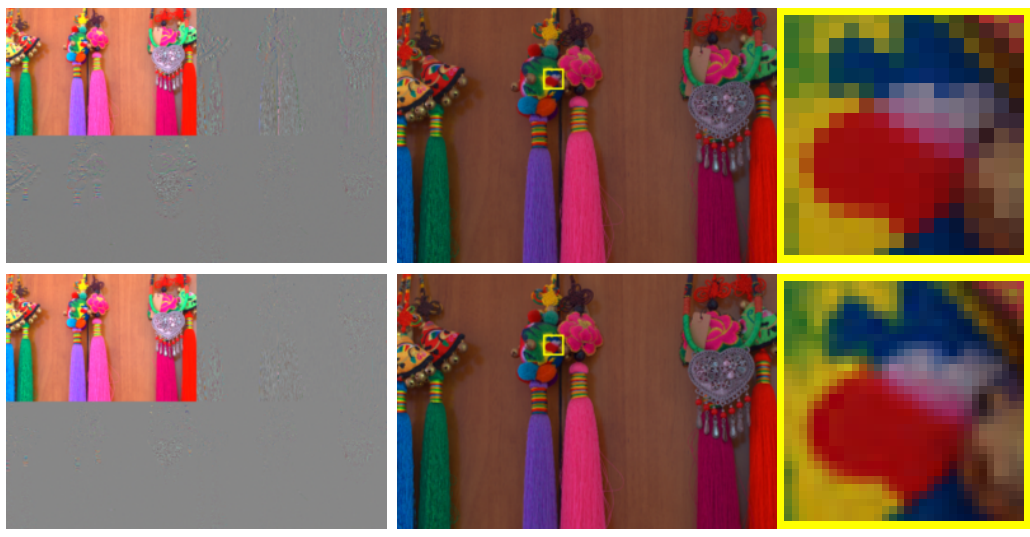

Fig. 1: Subband and reconstructed images with high-frequency components removed (*#179*): (top) Haar WT and (bottom) OWT (Bior4.4 as a representative configuration).

DiffLL nor Wave-Mamba can fully provide. DiffLL suffers from blocking artifacts caused by the non-overlapping Haar WT, as well as insufficient recovery of high-frequency structures buried in noise. Wave-Mamba, while effective at detail enhancement, tends to degrade under extremely dark conditions due to the limitations of its 2D-Mamba backbone.

To address these limitations, OWDiff integrates two complementary modules. An Overlapped Wavelet Transform (OWT) enforces cross-block structural continuity to suppress blocking artifacts. However, this process also attenuates high-frequency details, which are then compensated by a low-frequency-guided High-Frequency Enhance Block (HFEBlock). By combining these modules within the DiffLL framework, OWDiff improves both structural continuity and fine-detail reconstruction while retaining the robustness and efficiency of DiffLL.

#### 3.1.1 OWT

To address the blocking artifacts introduced by the Haar WT, we replace it with an *Overlapped* WT (OWT), which performs analysis using wider support filters. Unlike the $2 \times 2$ block-wise Haar WT, the OWT shares filter responses across neighboring pixels, preserving cross-block continuity and structurally preventing blocking boundaries without requiring any postprocessing.

In this study, we adopt the Bior2.2 and Bior4.4 WTs, whose analysis filters are summarized in Table 1.[†] Bior2.2 WT computes low- and high-frequency components from five and three samples, providing better energy concentration and more accurate detail extraction than the Haar WT.

[†]Bior2.2 and Bior4.4 WTs correspond to scaled versions of the 5/3- and 9/7-tap WTs used in JPEG 2000 lossless and lossy modes, respectively.

Similarly, Bior4.4 WT captures more low-frequency energy and achieves higher reconstruction fidelity by reducing reliance on high-frequency subbands. Figure 1 presents the subband images of the 1-level wavelet transform and the reconstructed images obtained by zeroing the high-frequency subbands. OWT (Bior4.4 in this case) eliminates block-edge discontinuities, whereas the Haar WT exhibits visible block artifacts.

While OWT effectively suppresses blocking artifacts, its wider support filters redistribute part of the high-frequency energy into the low-frequency component, which may attenuate fine structures near block boundaries. This limitation motivates the introduction of a complementary high-frequency refinement module.

#### 3.1.2 HFEBlock

In low-light images, high-frequency details such as edges and textures are often buried in noise, and the HFRM in DiffLL tends to oversmooth these structures due to its emphasis on denoising.

To compensate for the high-frequency attenuation introduced by OWT, OWDiff incorporates an HFEBlock that performs low-frequency-guided refinement. The HFEBlock uses enhanced low-frequency features as stable guidance to selectively refine and recover high-frequency components. This mechanism enables the restoration of sharper edges and more accurate textures that are partially attenuated by OWT's wider support filters. Since low-frequency enhancement is handled by the WCDM, OWDiff adopts a lightweight design with one HFEBlock per level, which is sufficient for effective detail restoration. While the HFEBlock substantially improves frequency-domain detail reconstruction, it does not address structural continuity, making it strictly complementary to OWT's structural role. Integrating both modules allows OWDiff to simultaneously maintain structural continuity and recover high-frequency details effectively.

### 3.2 Procedure

The overall pipeline of OWDiff with the 2-level decomposition is illustrated in Fig. 2 and consists of four stages: frequency decomposition via OWT, low-frequency enhancement via WCDM, high-frequency refinement via HFEBlock, and frequency reconstruction via IOWT.

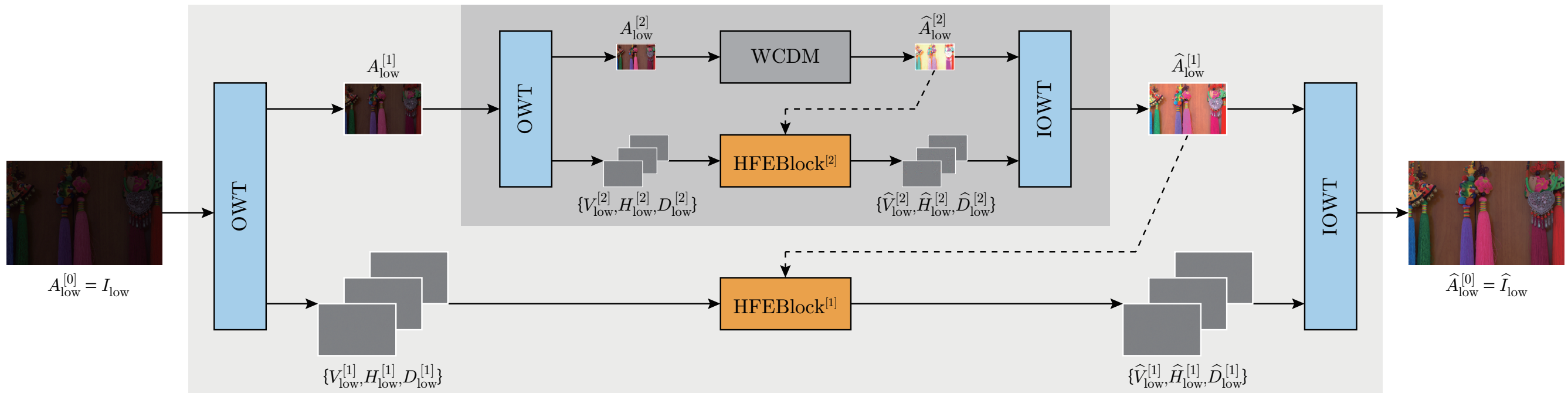

Fig. 2: Overall pipeline of our OWDiff with the 2-level decomposition.

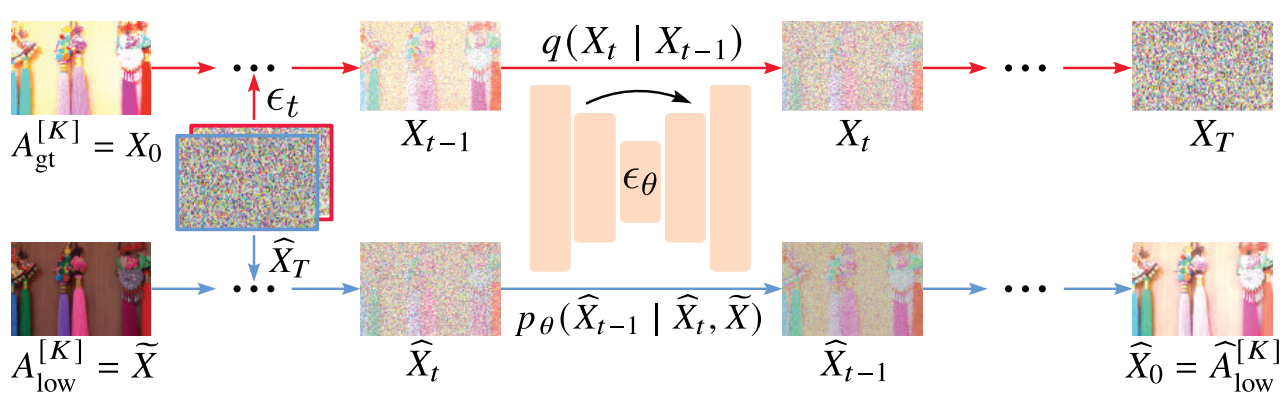

Fig. 3: Architecture of WCDM (red and blue arrows indicate the forward diffusion and denoising processes, respectively).

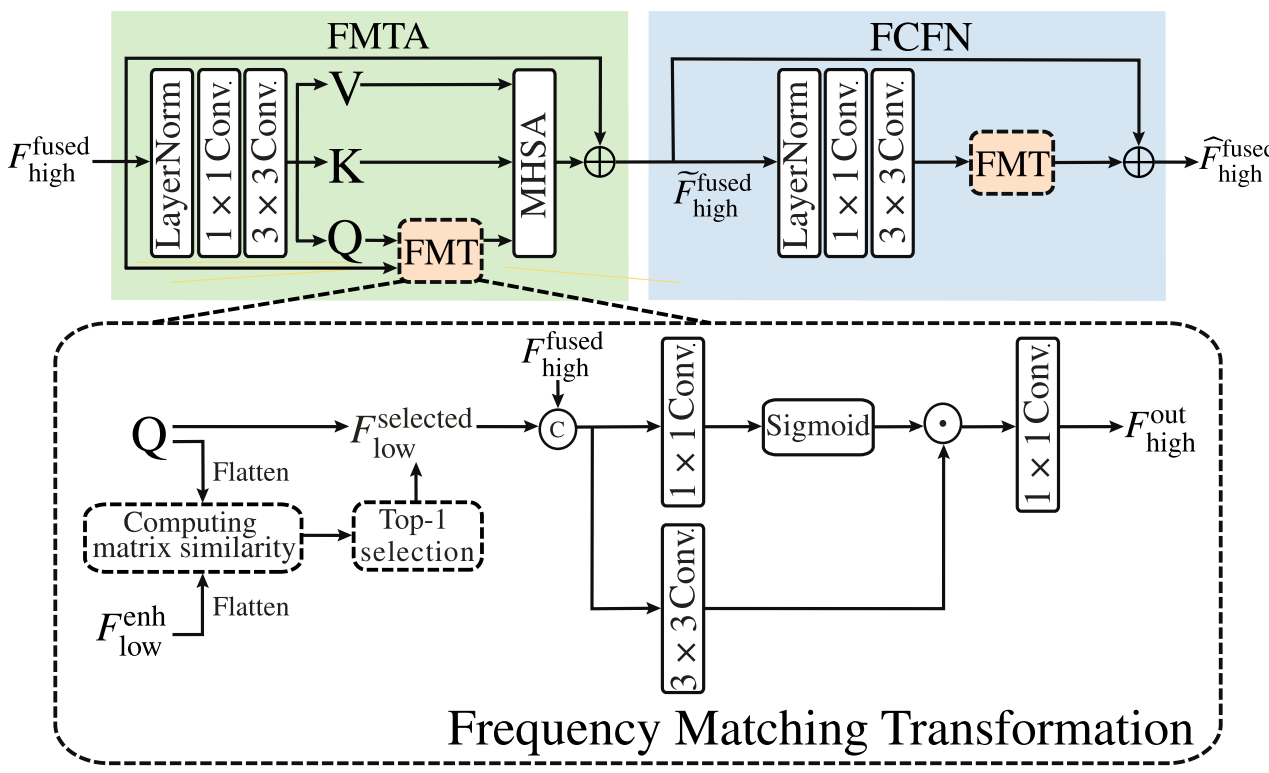

Fig. 4: Architecture of HFEBlock (© , ⊙, and MHSA indicate concatenation, Hadamard product, and Multi-Head Self-Attention, respectively).

### 3.2.1 Frequency Decomposition via OWT

A low-light image $I_{\text{low}} \in \mathbb{R}^{H\times W\times 3}$ is recursively decomposed into multiple frequency components using the OWT as

$$\{A_{\text{low}}^{[k]}, V_{\text{low}}^{[k]}, H_{\text{low}}^{[k]}, D_{\text{low}}^{[k]}\} = \text{OWT}(A_{\text{low}}^{[k-1]}), \tag{1}$$

where $A_{\text{low}}^{[k]}, V_{\text{low}}^{[k]}, H_{\text{low}}^{[k]}, D_{\text{low}}^{[k]} \in \mathbb{R}^{(H/2^k)\times(W/2^k)\times 3}$ denote the low-frequency subband, and the vertical, horizontal, and diagonal high-frequency subbands at a wavelet level $k \in \{1, 2, \dots, K\}$, respectively, and $A_{\text{low}}^{[0]} = I_{\text{low}}$. For supervised training, the normal-light (ground-truth) image $I_{\text{gt}}$ is decomposed in the same manner, producing $\{A_{\text{gt}}^{[k]}, V_{\text{gt}}^{[k]}, H_{\text{gt}}^{[k]}, D_{\text{gt}}^{[k]}\}$. Symmetric extension is applied at image boundaries to ensure structural continuity and suppress boundary artifacts during both decomposition and reconstruction.

### 3.2.2 Low-Frequency Enhancement via WCDM

As illustrated in Fig. 3, the WCDM enhances the lowest-frequency subband $A_{\text{low}}^{[K]}$ by performing a conditional diffusion-denoising process. During training, the forward diffusion process is applied to the ground-truth lowest-frequency component $A_{\text{gt}}^{[K]}$, while the lowest-frequency component $A_{\text{low}}^{[K]}$ is used as the conditioning input to the noise predictor $\epsilon_\theta$. In the denoising stage, $\epsilon_\theta$ predicts the noise for each diffusion step, generating the enhanced component $\widehat{A}_{\text{low}}^{[K]}$ by iteratively denoising a randomly initialized sample $\widehat{X}_T \in \mathbb{R}^{(H/2^K)\times(W/2^K)\times 3}$, where $t \in \{1, 2, \dots, T\}$ denotes the diffusion step. To stabilize training and enforce reconstruction consistency, the difference between $\widehat{A}_{\text{low}}^{[K]}$ and its target $A_{\text{gt}}^{[K]}$ is minimized. The final enhanced component $\widehat{A}_{\text{low}}^{[K]}$ is obtained by progressively denoising from $\widehat{X}_T$ to $\widehat{X}_0$.

### 3.2.3 High-Frequency Refinement via HFEBlock

As illustrated in Fig. 4, the HFEBlock follows the same internal structure as Wave-Mamba, consisting of the Frequency Matching Transformation Attention (FMTA) and the Frequency Correction Forward Network (FCFN). In OWDiff, only the input-output interfaces are adapted to the wavelet-domain representation.

The enhanced low-frequency subband $\widehat{A}_{\text{low}}^{[k]}$ and the high-frequency subbands $V_{\text{low}}^{[k]}$, $H_{\text{low}}^{[k]}$, and $D_{\text{low}}^{[k]}$ are first expanded from 3 channels to 48 channels through depth-wise separable convolutions. FMTA then performs channel-wise semantic alignment between the enhanced low-frequency features $F_{\text{low}}^{\text{enh}}$ and fused high-frequency features $F_{\text{high}}^{\text{fused}} \in \mathbb{R}^{(H/2^k)\times(W/2^k)\times 48}$. For each high-frequency channel, FMTA selects the most correlated low-frequency channel and constructs an aligned feature $F_{\text{low}}^{\text{selected}}$ as guidance. This process allows high-frequency components to be refined using structurally relevant low-frequency information, compensating for the high-frequency attenuation introduced by the OWT. The selected features $F_{\text{low}}^{\text{selected}}$ are concatenated with the fused high-frequency features $F_{\text{high}}^{\text{fused}}$, yielding the refined features $F_{\text{high}}^{\text{out}}$, which are further processed by the FCFN. Finally, a depth-wise separable convolution projects the output $\widehat{F}_{\text{high}}^{\text{fused}}$ back to 3 channels, producing the refined subbands $\widehat{V}_{\text{low}}^{[k]}$, $\widehat{H}_{\text{low}}^{[k]}$, and $\widehat{D}_{\text{low}}^{[k]}$, which preserve sharper edges and more reliable textures.

#### 3.2.4 Frequency Reconstruction via IOWT

The recovered $\widehat{A}_{\text{low}}^{[k]}$, $\widehat{V}_{\text{low}}^{[k]}$, $\widehat{H}_{\text{low}}^{[k]}$, and $\widehat{D}_{\text{low}}^{[k]}$ are transformed using the Inverse OWT (IOWT) as follows:

$$\widehat{A}_{\text{low}}^{[k-1]} = \text{IOWT}(\{\widehat{A}_{\text{low}}^{[k]}, \widehat{V}_{\text{low}}^{[k]}, \widehat{H}_{\text{low}}^{[k]}, \widehat{D}_{\text{low}}^{[k]}\}). \tag{2}$$

Here, $\widehat{A}_{\text{low}}^{[k-1]}$ corresponds to the final enhanced image $\widehat{I}_{\text{low}}$.

### 3.3 Loss Function

Following DiffLL, OWDiff is trained using a total loss

$$\mathcal{L}_{\text{total}} = \mathcal{L}_{\text{diff}} + \mathcal{L}_{\text{detail}} + \mathcal{L}_{\text{content}}, \tag{3}$$

which combines diffusion consistency, high-frequency accuracy, and image-level fidelity. First, the diffusion loss $\mathcal{L}_{\text{diff}}$ supervises the WCDM by minimizing both the noise prediction error and the reconstruction error of the lowest-frequency subband:

$$\begin{aligned}\mathcal{L}_{\text{diff}} = &\mathbb{E}_{X_0,t,\epsilon_t\sim\mathcal{N}(0,I)}\left[\|\epsilon_t - \epsilon_\theta(X_t, \widetilde{X}, t)\|_{\text{F}}^2\right] \\ &+ \|\widehat{A}_{\text{low}}^{[K]} - A_{\text{gt}}^{[K]}\|_{\text{F}}^2,\end{aligned} \tag{4}$$

where $X_t$ is the latent variable at step $t$, $\epsilon_t$ is the sampled noise, and $\widetilde{X} = A_{\text{low}}^{[K]}$ denotes the conditioning input. Next, the detail loss $\mathcal{L}_{\text{detail}}$ encourages accurate reconstruction of the high-frequency subbands and stabilizes them via total variation regularization:

$$\mathcal{L}_{\text{detail}} = \sum_{\substack{k\in\{1,2,\dots,K\}\\ S\in\{V,H,D\}}} \left(\lambda_1\|\widehat{S}_{\text{low}}^{[k]} - S_{\text{gt}}^{[k]}\|_{\text{F}}^2 + \lambda_2\text{TV}(\widehat{S}_{\text{low}}^{[k]})\right). \tag{5}$$

Here, $\lambda_1 = 0.1$ and $\lambda_2 = 0.01$ are weighting factors, and $\text{TV}(\cdot)$ denotes the total variation regularization term. Finally, the content loss $\mathcal{L}_{\text{content}}$ ensures global consistency between the enhanced output and the ground truth by combining the $\ell_1$ error and the structural similarity index:

$$\mathcal{L}_{\text{content}} = \|\widehat{I}_{\text{low}} - I_{\text{gt}}\|_1 + (1 - \text{SSIM}(\widehat{I}_{\text{low}}, I_{\text{gt}})). \tag{6}$$

Together, these three loss terms effectively guide OWDiff to improve brightness, preserve fine details, and maintain structural fidelity during training.

## 4. Experiments

### 4.1 Settings

The proposed model was trained using the LOLv1 [12] dataset, which comprises $1,000$ synthetic pairs and 485 real-world scene pairs. The remaining 15 real-world pairs from LOLv1 were reserved for evaluation. A total of 100 test pairs from the LOLv2-real [15] dataset were used to evaluate generalization capability. OWDiff was trained for $1,000$ epochs with 200 forward diffusion time steps, while 10 steps were used for the denoising process during both training and inference. All experiments and runtime measurements were conducted on a single NVIDIA RTX 5090 GPU. Inference time was measured by averaging the processing duration per image at the original $400 \times 600$ resolution without resizing. The initial learning rate was $1\times10^{-4}$ for $K = 1$ and $2\times10^{-4}$ for $K = 2$, with batch sizes of 4 and 16, respectively. The learning rate was decayed by a factor of 0.8 every $5 \times 10^3$ iterations. For comparison, we evaluated six existing methods: SNR-Aware [20], Retinexformer [21], GSAD [22], DiffLL [23], DiffUIR [24], and Wave-Mamba† [25]. For DiffUIR, we used the publicly available pretrained model. SNR-Aware, Retinexformer, GSAD, and Wave-Mamba were retrained under the same hyperparameter configuration used in OWDiff for $K = 1$, and DiffLL was retrained using exactly the same settings. Quantitative evaluation was conducted using the mean Peak Signal-to-Noise Ratio (PSNR) [dB], Structural Similarity Index Measure (SSIM), Learned Perceptual Image Patch Similarity (LPIPS), and inference time (Time) [seconds per image (spi)].

### 4.2 Evaluations

Figure 5 shows a qualitative comparison between OWDiff (Bior4.4, $K = 1$) and existing methods. The highlighted regions are selected for each image to illustrate representative failure cases such as blocking artifacts, unnatural textures, and over-enhancement, and the same regions are consistently applied across all methods for that image to ensure a fair and unbiased comparison. Across existing methods, several issues are observed, including misexposure, color distortion, and texture degradation. While GSAD may recover clearer details, it often suffers from over-enhancement and color shifts, resulting in unstable performance across different scenes. In contrast, methods such as SNR-Aware, Retinexformer, DiffUIR, and Wave-Mamba tend to produce unnatural or blurred textures in smooth or dark regions, while also weakening structural details. Meanwhile, DiffLL exhibits blocking artifacts caused by the Haar WT. By mitigating these issues, OWDiff achieves a better balance between artifact suppression and detail preservation, producing results without blocking artifacts or severe over-smoothing.

† We set $K = 3$ in accordance with the default configuration of the original paper, which provides the best trade-off between frequency decomposition depth and computational efficiency.

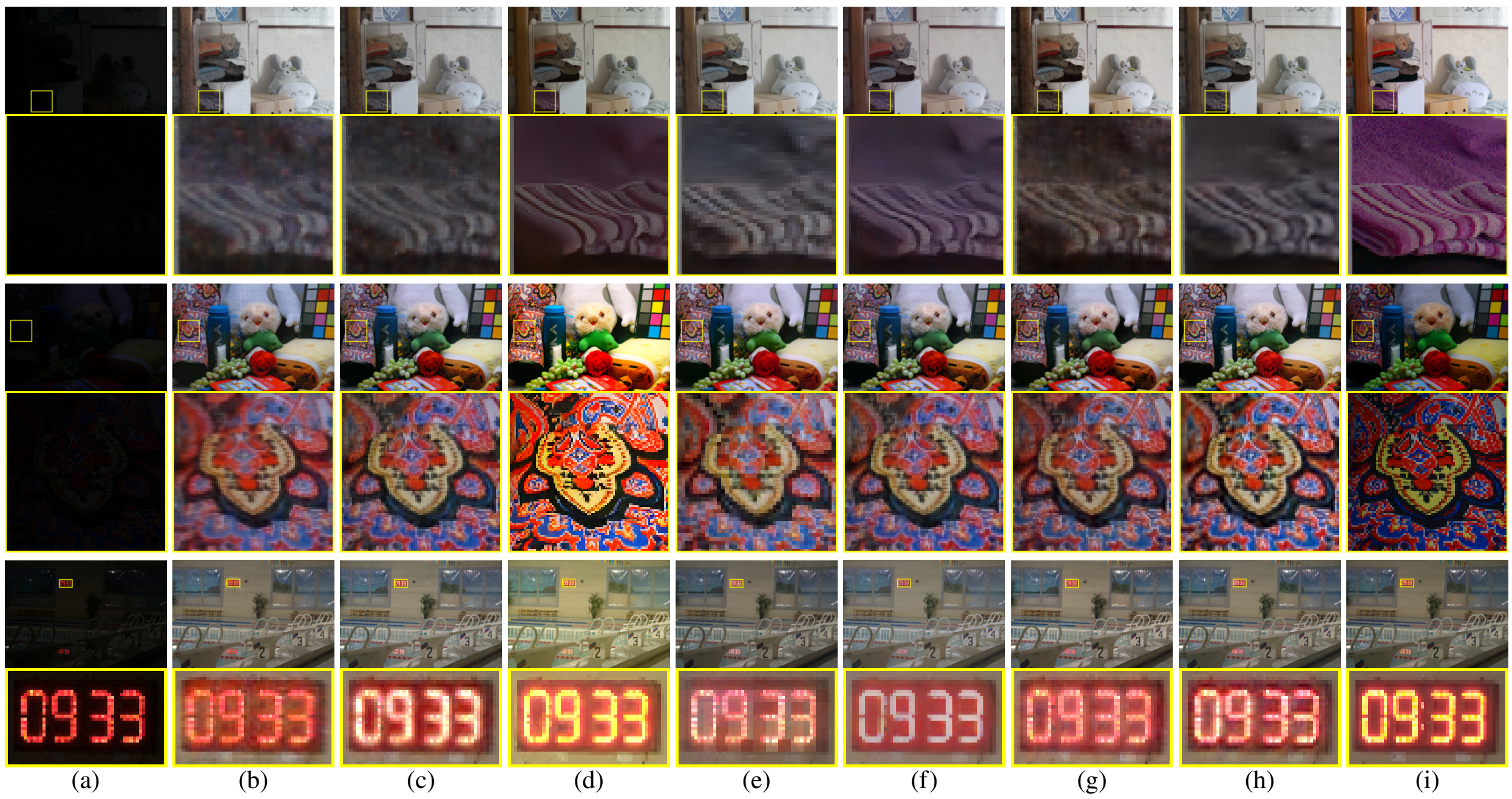

Fig. 5: Qualitative comparison among LLIE methods (*#23*, *#493*, and *#00747*): (a) input low-light images, (b) SNR-Aware [20], (c) Retinexformer [21], (d) GSAD [22], (e) DiffLL ($K = 1$) [23], (f) DiffUIR [24], (g) Wave-Mamba [25], (h) OWDiff (Bior4.4, $K = 1$), and (i) ground-truth.

Table 2: Quantitative comparisons on LOLv1 and LOLv2-real datasets.

| Method | LOLv1 | | | | LOLv2-real | | | |
|---|---|---|---|---|---|---|---|---|
| | PSNR [dB] ↑ | SSIM ↑ | LPIPS ↓ | Time [spi] ↓ | PSNR [dB] ↑ | SSIM ↑ | LPIPS ↓ | Time [spi] ↓ |
| SNR-Aware [20] | <u>23.19</u> | 0.800 | 0.344 | <u>0.026</u> | 24.17 | 0.832 | 0.286 | <u>0.030</u> |
| Retinexformer [21] | 21.81 | 0.808 | 0.360 | **0.023** | 26.92 | 0.860 | 0.212 | **0.025** |
| GSAD [22] | 22.32 | **0.848** | 0.357 | 0.425 | 24.77 | 0.891 | 0.260 | 0.438 |
| DiffLL ($K = 1$) [23] | 22.75 | 0.831 | 0.344 | 0.101 | 28.01 | 0.887 | 0.185 | 0.103 |
| DiffUIR [24] | 21.96 | 0.824 | 0.350 | 0.234 | 24.67 | 0.871 | 0.260 | 0.241 |
| Wave-Mamba [25] | 22.61 | 0.827 | 0.332 | 0.038 | 25.07 | 0.872 | 0.237 | 0.042 |
| OWDiff (Bior2.2, $K = 1$) | 23.11 | 0.844 | <u>0.322</u> | 0.104 | <u>28.28</u> | <u>0.897</u> | **0.172** | 0.105 |
| OWDiff (Bior4.4, $K = 1$) | **23.50** | <u>0.847</u> | **0.320** | 0.106 | **28.41** | **0.899** | <u>0.176</u> | 0.108 |

*The first and second rankings are bolded and underlined, respectively.

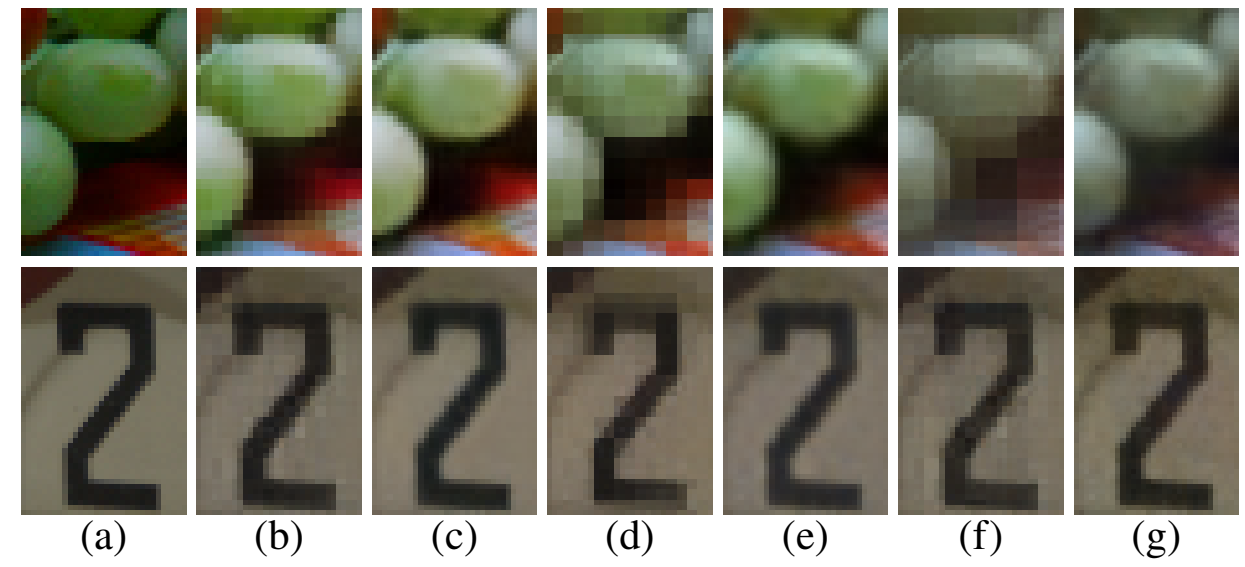

Fig. 6: Qualitative comparison of different wavelet levels (*#493 and #00747*): (a) ground-truth, (b) DiffLL [23] ($K = 1$), (c) OWDiff ($K = 1$), (d) DiffLL [23] ($K = 2$), (e) OWDiff ($K = 2$), (f) DiffLL [23] ($K = 3$), and (g) OWDiff ($K = 3$)

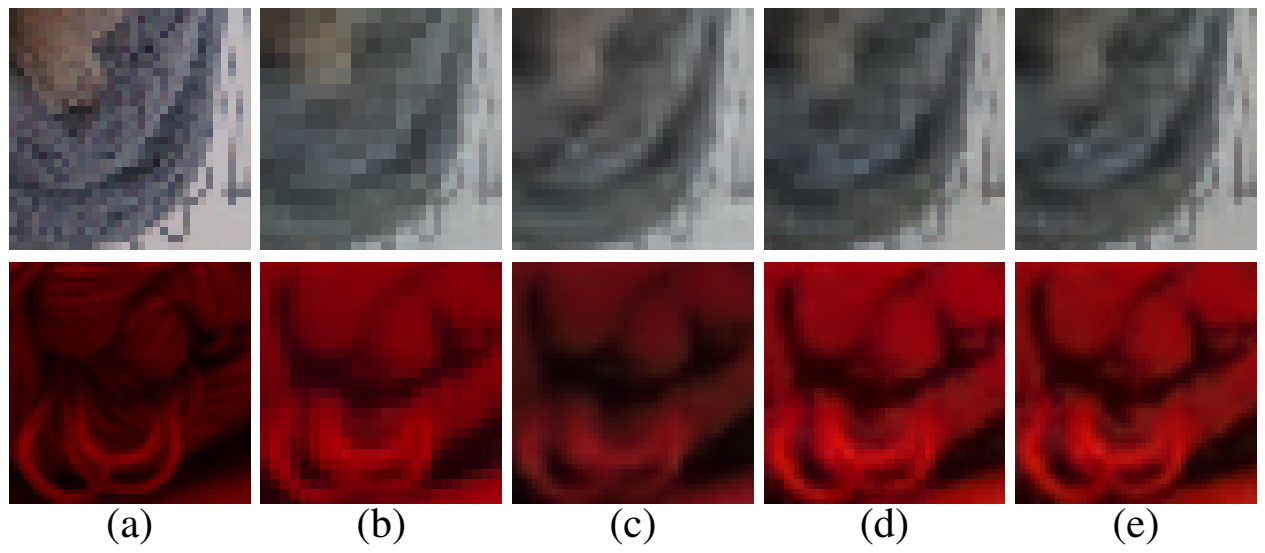

Fig. 7: Qualitative comparison highlighting the contributions of OWT and HFEBlock (*#23 and #493*): (a) ground-truth, (b) DiffLL [23], (c) DiffLL-OWT, (d) DiffLL-HFE, and (e) OWDiff.

Table 2 provides a quantitative comparison between OWDiff and existing methods. OWDiff (Bior4.4, $K = 1$) achieves a clear improvement over the original DiffLL (Haar, $K = 1$), increasing PSNR by 0.75 dB (23.50 vs. 22.75), and

Table 3: Quantitative comparison of different wavelet levels.

| Method | $K$ | PSNR[dB] ↑ | SSIM ↑ | LPIPS ↓ | Time [spi] ↓ |
|---|---|---|---|---|---|
| DiffLL [23] | 1 | 22.75 | <u>0.831</u> | 0.344 | 0.101 |
| | 2 | 22.35 | 0.813 | 0.354 | 0.053 |
| | 3 | 21.80 | 0.790 | 0.367 | **0.028** |
| OWDiff | 1 | **23.50** | **0.847** | **0.320** | 0.106 |
| | 2 | <u>23.00</u> | 0.828 | <u>0.339</u> | 0.056 |
| | 3 | 22.28 | 0.812 | 0.358 | <u>0.034</u> |

*The first and second rankings are bolded and underlined, respectively.

Table 4: Quantitative comparison highlighting the contributions of OWT and HFEBlock.

| Method | PSNR [dB] ↑ | SSIM ↑ | LPIPS ↓ | Time [spi] ↓ |
|---|---|---|---|---|
| DiffLL [23] | 22.75 | 0.831 | 0.344 | **0.101** |
| DiffLL-OWT | 23.00 | <u>0.843</u> | 0.335 | 0.104 |
| DiffLL-HFE | <u>23.47</u> | 0.841 | <u>0.329</u> | <u>0.102</u> |
| OWDiff | **23.50** | **0.847** | **0.320** | 0.106 |

*The first and second rankings are bolded and underlined, respectively.

further achieving higher SSIM and lower LPIPS, indicating superior structural fidelity and perceptual quality. As discussed in Section 3.1.1, although the use of Bior4.4 WT incurs a slightly higher inference time than Bior2.2 WT, OWDiff achieves better restoration performance across most metrics. For $K = 1$, OWDiff ranks among the top two methods across all metrics, demonstrating the effectiveness of integrating the OWT and the HFEBlock in preventing blocking artifacts and accurately recovering high-frequency structures. In addition, OWDiff achieves faster inference than recent diffusion-based models such as GSAD and DiffUIR, while maintaining computational efficiency.

### 4.3 Ablation Study

We performed two ablation studies on the LOLv1 dataset to examine the key design choices of OWDiff. The first evaluated the effect of the wavelet level by comparing $K =$ 1, 2, and 3 using the Bior4.4 WT as the OWT. The second assessed the contributions of the two main components: the OWT (DiffLL-OWT) and the HFEBlock (DiffLL-HFE), with the OWT set to the Bior4.4 WT with $K = 1$.

Table 3 presents a quantitative comparison among $K =$ 1, 2, and 3 for both DiffLL and OWDiff. While $K = 1$ consistently achieves the best reconstruction quality, $K = 2$ and 3 offer improved computational efficiency, indicating that the proposed design scales favorably with increasing wavelet depth. Figure 6 qualitatively demonstrates that $K = 1$ produces clearer and more natural details than those of higher levels. As $K$ increases, the reduced spatial resolution limits the capacity of the WCDM, leading to structural degradation and reduced color fidelity. Reduced color fidelity is observed in both DiffLL and OWDiff, whereas the nature of structural degradation differs between the two methods. Consequently, DiffLL exhibits noticeable blocking artifacts due to its non-overlapping transform, along with noise amplification in smooth regions. In contrast, OWDiff effectively suppresses such artifacts but tends to produce slightly over-smoothed results at higher $K$, as the WCDM receives less informative guidance for accurate restoration.

Table 4 quantitatively evaluates the contribution of each component. Removing the OWT led to a clear degradation across all metrics, indicating degraded reconstruction quality. Notably, although an additional HFEBlock is introduced, the inference time at $K = 2$ is approximately half that of $K = 1$. This is because the dominant computational cost lies in the WCDM, whose cost is reduced at higher wavelet levels due to lower spatial resolution. In addition, the ablation study for the HFEBlock showed consistent improvements, demonstrating the effectiveness of low-frequency guidance. These quantitative results confirmed that both the OWT and the HFEBlock are essential for achieving the full performance of OWDiff. Figure 7 further reveals that removing the OWT reintroduces visible blocking artifacts, whereas the full OWDiff successfully achieves blocking artifact-free enhancement with sharper edges and more reliable textures.

Finally, due to space limitations, the experimental results on LOLv2-real are omitted; however, consistent trends were confirmed across all ablation experiments, in line with those observed on LOLv1.

## 5. Conclusion

In this paper, we presented Overlapped Wavelet Diffusion (OWDiff), a Low-Light Image Enhancement (LLIE) framework that integrates two complementary improvements into the original DiffLL model. The first is the replacement of the Haar Wavelet Transform (WT) with an Overlapped WT (OWT), which preserves structural continuity across neighboring regions and eliminates the blocking artifacts inherent to non-overlapping block transforms. The second is the introduction of a low-frequency-guided High-Frequency Enhance Block (HFEBlock), which compensates for the high-frequency attenuation introduced by OWT and restores edges and textures that would otherwise remain blurred or overly smoothed. By combining OWT for structural consistency and HFEBlock for detail recovery, OWDiff achieves both blocking artifact-free reconstruction and high-fidelity enhancement. Experiments demonstrated that OWDiff provides superior quantitative accuracy and perceptual quality compared with existing LLIE methods, while maintaining practical computational efficiency.

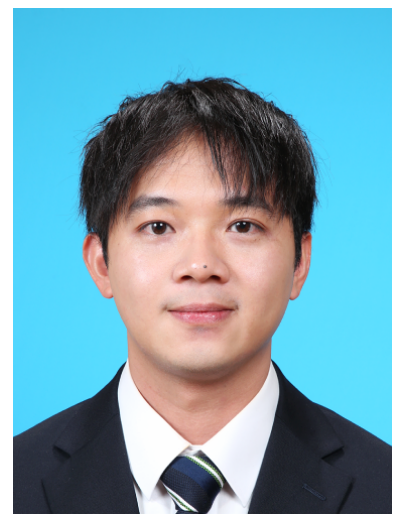

**Fen Peng** received the B.E. degree in Radio and TV Engineering from Communication University of China, China, in 2018. From 2018 to 2019, he was with Guangdong Radio and Television, China. From 2019 to 2023, he was with Zhuhai Media Group Co., Ltd., China. In 2024, he joined Master's Program in Computer Science, University of Tsukuba, Japan. His research interests include signal processing, computational photography, and computer vision.

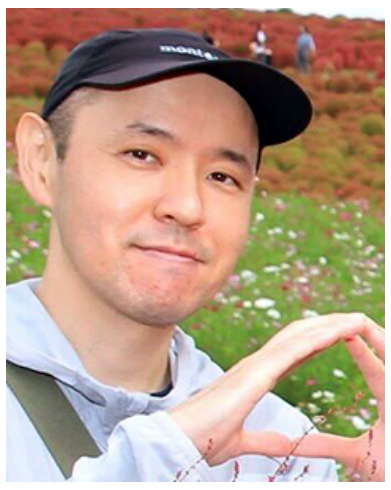

**Taizo Suzuki** received the B.E., M.E., and Ph.D. degrees in Electrical Engineering from Keio University, Japan, in 2004, 2006, and 2010, respectively. From 2006 to 2008, he was with Toppan Printing Co., Ltd., Japan. From 2008 to 2011, he was a Research Associate with the Global Center of Excellence (G-COE) Program at Keio University, Japan. From 2010 to 2011, he was a Research Fellow of the Japan Society for the Promotion of Science (JSPS) and a Visiting Scholar with the Video Processing Group, University of California, San Diego, La Jolla, CA, USA. From 2011 to 2012, he was an Assistant Professor with Nihon University, Japan. From 2024 to 2026, he was a Science and Technology Policy Fellow with the Council for Science, Technology and Innovation (CSTI), Cabinet Office, Government of Japan. In 2012, he joined the University of Tsukuba, Japan, as an Assistant Professor, where he has been an Associate Professor since 2019. His research interests include signal processing, machine learning, and their applications to image and video content. He served as an Associate Editor (2017–2021) and an Area Editor (2023–2025) for the *IEICE Transactions on Fundamentals of Electronics, Communications and Computer Sciences*.

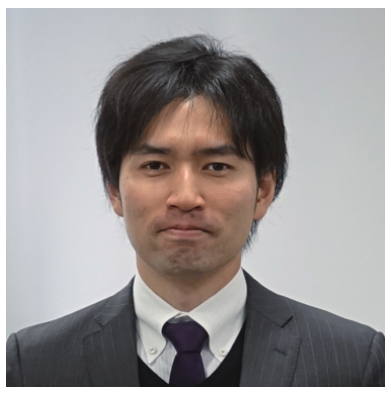

**Seisuke Kyochi** received the B.S. degree in mathematics from Rikkyo University, Japan, in 2005, and the M.E. and Ph.D. degrees from Keio University, Yokohama, Japan, in 2007 and 2010, respectively. From 2010 to 2012, he was a Researcher at NTT Cyberspace Laboratories. From 2012 to 2015, he was a Lecturer with the Faculty of Environmental Engineering, The University of Kitakyushu, Japan, where he was an Associate Professor from 2015 to 2021. In 2021, he joined Kogakuin University, Japan, as an Associate Professor. In 2026, he joined Hosei University, Japan, where he is currently a Professor. His research interests include convex optimization for signal recovery and the theory and design of wavelets and filter banks for signal processing applications.